\documentclass{article}
\pdfoutput=1
\usepackage{vmargin}
\setpapersize{A4}
\usepackage{graphics}
\usepackage{epsfig}
\usepackage{setspace}
\usepackage{algorithm}
\usepackage{algorithmic}
\usepackage{theapa}
\usepackage{lscape}

\usepackage{ifpdf}
\usepackage{color}
\ifpdf
\else

\fi

\usepackage{url}

\newcommand{\jeromecut}[1]{}%
\newcommand{\kamelcut}[1]{}%

\date{}

\title{Data Mining-based Materialized View and Index Selection \\in Data Warehouses}

\author{Kamel Aouiche and J\'{e}r\^{o}me Darmont\\
University of Lyon (ERIC Lyon 2)\\
5 avenue Pierre Mend\`{e}s-France\\
69676 Bron Cedex\\
France\\
E-mail: firstname.lastname@eric.univ-lyon2.fr}

\begin{document}
\maketitle

\doublespacing

\abstract{Materialized views and indexes are physical structures for accelerating data
access that are casually used in data warehouses. However, these data structures generate some
maintenance overhead. They also share the same storage space. Most existing
studies about materialized view and index selection consider these structures
separately. In this paper, we adopt the opposite stance and couple
materialized view and index selection to take view-index interactions into account
and achieve efficient storage space sharing. Candidate materialized views and indexes are
selected through a data mining process. We also exploit cost models that
evaluate the respective benefit of indexing and view materialization, and help select a relevant
configuration of indexes and materialized views among the candidates. Experimental results show that
our strategy performs better than an independent selection of
materialized views and indexes.

}




\section{Introduction}

Large-scale usage of databases in general and data warehouses in particular
requires an administrator whose principal role is data management, both at the
logical level (schema definition) and physical level (files and disk storage),
as well as performance optimization. With the wide development of Database
Management Systems (DBMSs), minimizing the administration function has become
crucial~\cite{cha97eff}. One important administration task is the selection of suitable physical
structures to improve system performance by minimizing data access
time~\cite{fin88phy}.

Among techniques adopted in data warehouse relational implementations for
improving query performance, view materialization and indexing are presumably
the most effective~\cite{vie03riz}. Materialized views are physical
structures that improve data access time by precomputing intermediary results.
Therefore, end-user queries can be efficiently processed through data stored
in views and do not need to access the original data. Indexes are also physical
structures that allow direct data access. They avoid sequential scans
and thereby reduce query response time. Nevertheless, exploiting either
materialized views or indexes requires additional storage space and entails
maintenance overhead when refreshing the data warehouse. The issue is thus to select
an appropriate configuration (set) of materialized views and indexes
that minimizes query response time and the selected data structures' maintenance cost, given a limited storage space.

The literature regarding materialized view and index selection in relational
databases and data warehouses is quite abundant. However, we have identified
two key issues requiring enhancements. First, the actual selection of suitable
candidate materialized views and indexes is rarely addressed in existing
approaches. Most of them indeed present scaling problems at this level. Second,
none of these approaches takes into account the interactions that may exist
between materialized views, between indexes, and between indexes and
materialized views (including the approaches that simultaneously select both
materialized views and indexes).


In this paper, we present a novel strategy for optimizing data  warehouse
performance that aims at addressing both these issues. We have indeed designed a
generic approach whose objective is to automatically propose solutions to data
warehouse administrators for optimizing data access time. The principle of
this approach is to apply data mining techniques on a workload (set of queries)
that is representative of data warehouse usage in order to deduce a
quasi-optimal configuration of materialized views and/or indexes. Data mining
actually helps reduce the selection problem's complexity and
improves scalability. Then, cost models help select among the selected materialized views and indexes
the most efficient in terms of performance gain/overhead ratio. We have applied
our approach on three related problems: isolate
materialized view selection, isolate index selection and joint materialized view and index selection. In
the last case, we included index-view interactions in our cost models.

The remainder of this paper is organized  as follows.
Section~\ref{sec:RelatedWork} presents and discusses the state of the art
regarding materialized view and index selection.
Section~\ref{sec:DataMiningBasedMaterializedViewAndIndexSelection} motivates
and presents the principle of our performance optimization approach.
Section~\ref{sec:Applications} further details how we apply this approach to
 isolate materialized view selection, isolate index selection and joint
materialized view and index selection, respectively. We particularly focus on
joint materialized view and index selection, which is our latest development.
Section~\ref{sec:Experiments} presents the experimental results we achieved to
illustrate our approach's relevance. Finally, we conclude this paper and
provide research perspectives in Section~\ref{sec:ConclusionAndPerspectives}.

\section{Related work}
\label{sec:RelatedWork}

In this section, we first formalize the materialized view and index selection
problem, and then detail and discuss the state of the art regarding
materialized view selection, index selection and joint index and materialized
view selection, respectively.

\subsection{Materialized view and index selection: formal problem definition}

The materialized view and index selection problem consists  in building a set
of materialized views and indexes that optimizes the execution cost of a given
workload. This optimization may be realized under constraints, typically the
storage space available for storing these physical data structures.

Let $V_C$ and $I_C$ be two sets of materialized views and indexes, respectively,
that are termed candidate and are susceptible to reduce the execution cost of a
given query set $Q$ (generally supposed representative of system workload). Let
$O_C = V_C \cup I_C$. Let $S$ be the storage space allotted by the data
warehouse administrator to build objects (materialized views or indexes) from
set $O_C$. The joint materialized view and index selection problem consists in
building an object configuration $O \subseteq O_C$ that minimizes the execution
cost of $Q$, under storage space constraint. This NP-hard problem
\cite{com78dif,gup99sel-a} may be formalized as follows:
\begin{itemize}
    \item $cost(Q, O) = min(cost(Q, \vartheta)) ~\forall ~\vartheta \subseteq O_C$~;
    \item $\displaystyle{\sum_{o \in O}} size(o) \leq S$, where $size(o)$ is the disk space occupied by object $o$.
\end{itemize}

\subsection{Materialized view selection}

The materialized view selection problem has received significant attention in the literature. Related researches differ in several points:
\begin{enumerate}
    \item the way the set of candidate views $V_C$ is determined;
    \item the framework used to capture relationships between candidate views;
    \item the use of mathematical cost models vs. calls to the system's query optimizer;
    \item view selection in the relational or multidimensional context;
    \item multiple or simple query optimization;
    \item theoretical or technical solutions.
\end{enumerate}

Classical papers in materialized view selection introduce a lattice framework
that models and captures dependency (ancestor or descendent) among aggregate
views in a multidimensional context~\cite{har96imp,bar97mat,kot99dyn,uch99pro}.
This lattice is greedily browsed with the help of cost models to select the
best views to materialize. This problem has first been addressed in one data
cube and then extended to multiple cubes~\cite{shu00mat}. Another theoretical
framework, the AND-OR view graph, may also be used to capture the relationships
between views~\cite{goe99des,nad02ach,val02vie,gup05sel}. Unfortunately, the majority of these
solutions are theoretical and are not truly scalable.


A wavelet framework for adaptively representing multidimensional data cubes has
also been proposed~\cite{smi04wav}. This method decomposes data cubes into an
indexed hierarchy of wavelet view elements that correspond to partial and
residual aggregations of data cubes. An algorithm greedily selects a
non-expensive set of wavelet view elements that minimizes the average
processing cost of data cube queries. In the same spirit, \citeA{sis02dwa}
proposed the Dwarf structure, which compresses data cubes. Dwarf identifies
prefix and suffix redundancies within cube cells and factors them out by
coalescing their storage. Suppressing redundancy improves the maintenance and
interrogation costs of data cubes. These approaches are very interesting, but
they are mainly focused on computing efficient data cubes by changing their
physical design.


Other approaches detect common sub-expressions within workload queries in the
relational context~\cite{gol01opt,bar03sel,vie03riz}. The view selection
problem then consists in finding common subexpressions corresponding to
intermediary results that are suitable to materialize. However, browsing is
very costly and these methods are not truly scalable with respect to the number
of queries.

Finally, the most recent approaches are workload-driven. They syntactically
analyze a workload to enumerate relevant candidate views~\cite{agr00aut}. By
exploiting the system's query optimizer, they greedily build a configuration of
the most pertinent views. A workload is indeed a good starting point to predict
future queries because these queries are probably within or syntactically close
to a previous query workload. In addition, extracting candidate views from the
workload ensures that future materialized views will probably be used when
processing queries.

\subsection{Index selection}
\label{soa-index}

The index selection problem has been studied for many years in
databases~\cite{fin88phy,fra92ada,agr00aut,val00db2,fel03nea,kra03gen,cha04ind}.
In the more specific context of data warehouses, existing research studies may
be clustered into two families: algorithms that optimize maintenance
cost~\cite{lab97phy} and algorithms that optimize query response
time~\cite{gup97ind,agr01mat,gol02ind}. In both cases, optimization is realized
under storage space constraint. In this paper, we focus on the second family of
solutions, which is relevant in our context. Studies falling in this category
may be further categorized depending on how the set of candidate indexes $I_C$
and the final configuration of indexes $I$ are built.

Selecting a set of candidate indexes may be automatic or manual. Warehouse
administrators may indeed appeal to their expertise and manually provide, from
a given workload, a set of candidate indexes~\cite{fra92ada,cho93ind,cho93sel}.
Such a choice is however subjective. Moreover, the task may be very hard to
achieve when the number of queries is very high. In opposition, candidate
indexes can also be extracted automatically, through a syntactic analysis of
queries~\cite{cha97eff,val00db2,gol02ind}. Such an analysis depends on the
DBMS, since each DBMS is queried through a specific syntax derived from the SQL
standard.

The methods for building a final index configuration from candidate indexes may be categorized into:
\begin{enumerate}
    \item ascending or descending greedy methods;
    \item methods derived from genetic algorithms;
    \item methods assimilating the selection problem to the well-known knapsack optimization problem.
\end{enumerate}


Ascending greedy methods start from an empty set of candidate
indexes~\cite{wha87ind,fra92ada,cho93sel,cha97eff}. They incrementally add in
indexes minimizing cost. This process stops when cost ceases decreasing.
Contrarily, descending greedy methods consider the whole set of candidate
indexes as a starting point. Then, at each iteration, indexes are
pruned~\cite{wha87ind,cho93ind}. If workload cost before pruning is lower
(respectively, greater) than workload cost after pruning, the pruned indexes
are useless (respectively, useful) for reducing cost. The pruning process stops
when cost increases after pruning.

Genetic algorithms are commonly used to resolve optimization problems. They
have been adapted to the index selection problem~\cite{kra03gen}. The initial
population is a set of input indexes (an index is assimilated to an
individual). The objective function to optimize is the workload cost
corresponding to an index configuration. The combinatory construction of an
index configuration is realized through the crossover, mutation and selection
genetic operators. Eventually, the index selection problem has also been
formulated in several studies as a knapsack
problem~\cite{ip83sel,gun99opt,val00db2,fel03nea} where indexes are objects,
index storage costs represent object weights, workload cost is the benefit
function, and storage space is knapsack size.


\subsection{Joint materialized view and index selection}

Few research studies deal with simultaneous index and materialized view
selection. \citeA{agr00aut} have proposed three strategies. The first one,
MVFIRST, selects materialized views first, and then indexes, taking the
presence of selected views into account. The second alternative, INDFIRST,
selects indexes first, and then materialized views. The third alternative,
joint enumeration, processes indexes, materialized views and indexes over these
views at the same time. According to the authors, this approach is more
efficient than MVFIRST and INDFIRST, but no further details are provided.

\citeA{bel00eff} studied storage space distribution among materialized views
and indexes. First, a set of materialized views and indexes is designed as an
initial solution. Then, the approach iteratively reconsiders the solution to
further reduce execution cost, by redistributing storage space between indexes
and materialized views. Two agents are in perpetual competition: the index spy
(respectively, view spy) steals some space allotted to materialized views
(respectively, indexes), and \emph{vice versa}. The recovered space is used to
create new indexes (respectively, materialized views) and prune views
(respectively, indexes), according to predefined replacement policies.

Another approach \textit{a priori} determines a trade-off between storage space
allotted to indexes and materialized views, depending on  query
definition~\cite{vie03riz}. According to the authors, the key factors to
leverage query optimization is aggregation level, defined by the attribute list
of \texttt{Group by} clauses in SQL queries, and the selectivity of attributes
present in \texttt{Where} and \texttt{Having} clauses. View materialization
indeed provides a great benefit for queries involving coarse granularity
aggregations (few attributes in the \texttt{Group by} clause) because they
produce few groups among a large number of tuples. On the other hand, indexes
provide their best benefit with queries containing high selectivity attributes.
Thus, queries with fine aggregations and high selectivity stimulate indexing,
while queries with coarse aggregations and weak selectivity encourage view
materialization.


Finally, \citeA{bruno06} have recently worked on refining the physical design
of relational databases. Their objective was to automatically improve an
expert's physical design, to take into account primordial constraints it might
violate. Hence, they proposed a transformation architecture base on two fusion
and reduction primitives that helps process indexes and materialized views in a
unified way.

\subsection{Discussion}
\label{sec:Discussion}

Existing studies related to index and materialized view selection are numerous
and diverse in the field of databases, and quite developed in the field of data
warehouses as well. However, we have identified two main points that could be
improved in these approaches.

\subsubsection{Candidate object selection}

Selecting candidate objects (materialized views and indexes) is rarely the
focus of existing approaches, most of which do not scale up well at this level.
Many index selection strategies indeed rest on human expertise (the warehouse
administrator's) to propose an initial candidate index configuration. Given the
size and complexity of most data warehouses, an automatic approach is mandatory
to apply these methods on a real-life scale. The most recent studies actually
take this option, by building the initial index configuration from system
workload.

With respect to materialized views, various data structures have been proposed
(lattices, graphes, wavelets...) to model inter-view relationships. None of
them scale up very well. For instance, browsing a candidate view lattice is
very costly when the input data cube is very large. Similarly, building view
graphs is as complex as the input workload is large. Hence, it is necessary to
carefully evaluate a strategy's complexity before adopting it, and to optimize
any data structure used.

\subsubsection{Inter-object interaction management}

None of the approaches we have presented in this section takes into account the
interactions that may exist between indexes, between materialized views, and
between index and views, including joint selection methods. Existing studies,
especially those assimilating the selection problem to the knapsack problem or
exploiting genetic algorithms, indeed compute the cost or benefit of an object
(index or materialized view) once only, before injecting it in their algorithm.
However, the relevance of selecting a given object may vary from one iteration
to the other if another, previously selected object interacts with the first
one. It is thus primordial to recompute costs or benefits dynamically before
object selection.

The nearest solution is the one by \citeA{bel00eff}. However, its object
replacement policies in the disk spaces allotted to indexes and materialized
views do not truly reflect index-view interactions. They indeed only consider
joint usage frequency in queries, and not the benefit an object brings with
respect to other objects.

\section{Data mining-based warehouse performance optimization approach}
\label{sec:DataMiningBasedMaterializedViewAndIndexSelection}

In this section, we first motivate our performance optimization approach. Then,
we present its general principle, detail how candidate objects are selected and
how a final object (materialized view and index) configuration is generated.

\subsection{Motivation}
\label{sec:Motivation}

In this paper, our objective is to address the issues identified in
Section~\ref{sec:Discussion}. First, to ensure that candidate object
(materialized view or index) selection scales up, it is necessary to devise an
automatic approach. Generally, this is achieved by syntactically analyzing the
system's query workload, which helps identify query attributes that might
support indexes or materialized views. These attributes are then systematically
combined to propose multi-attribute indexes or exhaustive view graphs. However,
this strategy later leads, in the selection phase, to consider irrelevant
objets, i.e., objects that do belong to the workload, but are not interesting
in the scope of indexing or view materialization.

To \emph{a priori} eliminate these irrelevant objects, we propose to exploit
data mining techniques to directly extract from the workload a configuration of
pertinent candidate objects. Our idea is to discover co-occurencies and
similarities between workload objects. For indexing, we base our approach on
the intuition that the importance of an attribute to index is strongly
correlated with its appearance frequency in the workload. For view
materialization, devising similar classes of queries also helps build views that
are likely to answer all the queries from a given class.

On the basis of the smallest possible set of candidate, all relevant objects,
we must then exploit an optimization algorithm (typically a greedy, knapsack or
genetic algorithm) to build a quasi-optimal object configuration. However, to
take index-view interactions into account, such algorithms must be modified. In
a given iteration, an object's cost indeed depends on previously selected
objects. Thus, it must be recomputed at each step. For simplicity reasons, we
implemented this approach in a greedy algorithm.

\subsection{General principle of our approach}
\label{sec:GeneralPrinciple}

\begin{figure}[hbt]
    \centering
    \includegraphics[width=9cm]{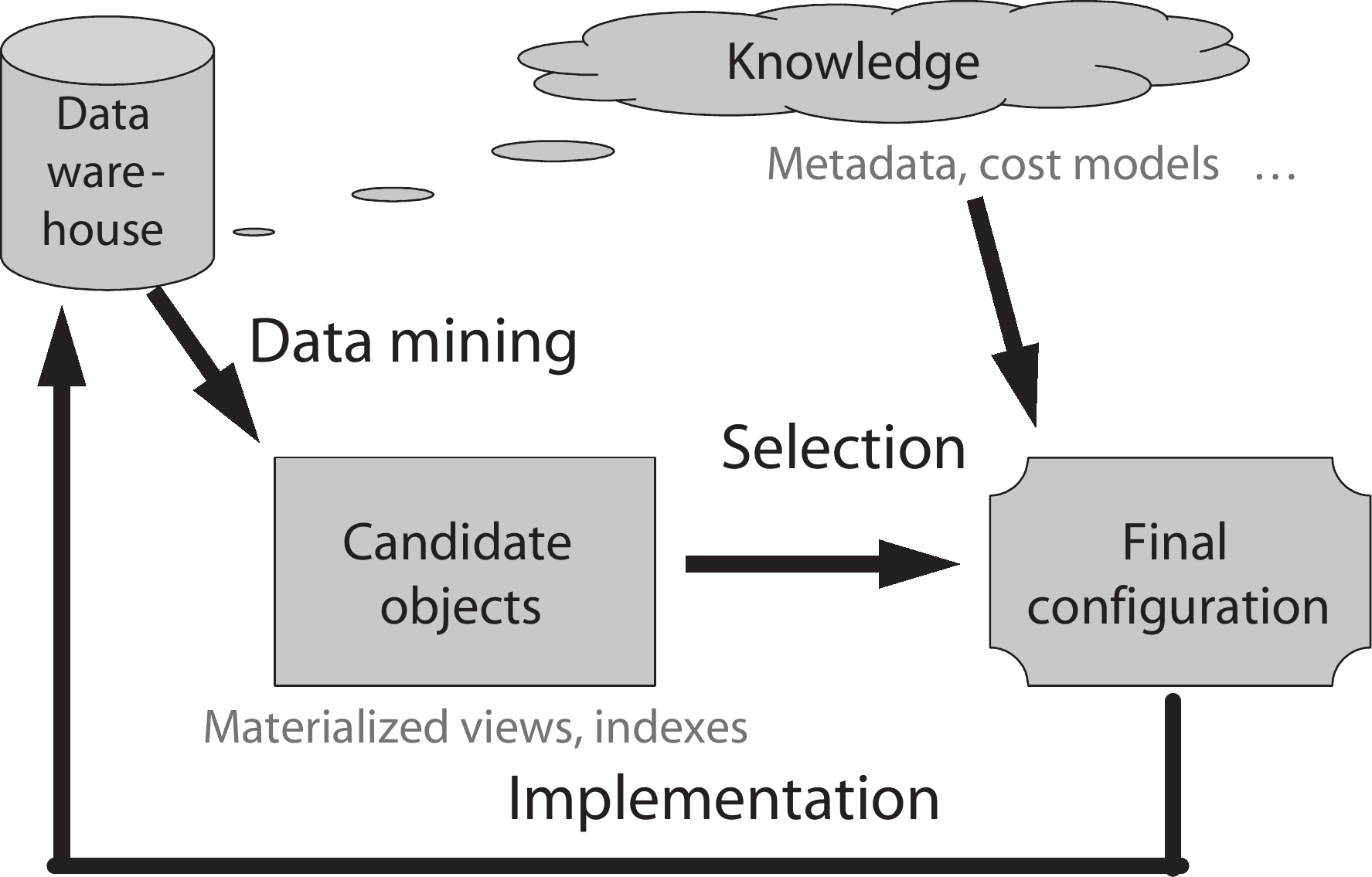}
    \caption{Principle of our automatic performance optimization approach}
    \label{principe_approche}
\end{figure}

Our automatic warehouse performance optimization approach
(Figure~\ref{principe_approche}) is not only based on information extracted
from the warehouse's data (statistics such as attribute selectivity, for
instance) or workload, but also on knowledge. This knowledge includes classical
warehouse metadata (we notably exploit the database schema), as well as
administration expertise, formalized in cost models (benefit induced by an
index and maintenance cost, for instance) or rules. Our approach proceeds in
two main steps, which are both piloted by knowledge.

The first step is building a candidate object configuration $O_C$. It consists
in syntactically analyzing the input workload, which helps identify attributes
that might be useful for view materialization or indexing. Applying rules
issued from administration-related knowledge can already reduce the size of
this attribute set. For instance, a low selectivity attribute such as gender,
which has only two values, is not a good candidate index. This set of
attributes is then structured as an attribute-value table that can be processed
by a data mining algorithm. The output of such an algorithm is directly the
candidate object configuration.

Since disk space is constrained, it would be impossible to exploit all the
candidate objects from $O_C$. Thus, the second step in our process is greedily
selecting a final object configuration $O$ from $O_C$. This algorithm exploits
cost models we have developed to express, e.g., the benefit brought by a
materialized view or an index, as well as their storage and maintenance costs
(Section~\ref{sec:Applications}). Eventually, the last step in our approach
consists in implementing the final object configuration in the data warehouse.

Note that we have designed this approach in a modular fashion, so that it is as
generic as possible. Completing the two main steps indeed brought us to perform
choices, but other options would be easy to consider. For instance, the data
mining technique we selected for building a candidate index configuration is
frequent itemset mining, but another study explored clustering instead
\cite{zaman04}. Besides, we have also used clustering for materialized view
selection. Similarly, other optimization algorithms could be substituted to the
greedy strategy we adopted to build the final object configuration. Our cost
models might also be easily replaced with others if necessary, or by calls to a
query optimizer, if it is accessible on the host DBMS.

\subsection{Candidate object selection}
\label{sec:CandidateObjectsSelection}

System workload is typically accessible from the host DBMS' transaction log. A
given workload is supposed representative if it has been measured during a time
period the warehouse administrator judges sufficient to anticipate upcoming
transactions.

Since we are more particularly interested in decision-support query performance
and not warehouse maintenance, we only consider interrogation query workloads
in this paper. These queries are typically composed of join operations between
the fact table and dimensions, restriction predicates and aggregation and
grouping operations. More formally, an analytic query $q$ may be expressed as
follows in relational algebra: $q = \Pi_{G,M} (\sigma_R (F \bowtie D_1 \bowtie
D_2 \bowtie ... \bowtie D_d))$, where $G$ is the set of attributes from
dimensions $D_i$ that are present in $q$'s grouping clause, $M$ is a set of
aggregate measures from fact table $F$ and $R$ a conjunction of predicates over
dimension attributes.

Attributes that may support materialized views or indexes belong to sets $G$
and $R$ \cite{cha97eff,gol02ind,val00db2,fel03nea}. We reference them in a
so-called ``query-attribute'' binary matrix whose rows represent workload
queries $q_i$ and whose columns are representative attributes $a_j$. The
general term $m_{ij}$ of this matrix is equal to one if attribute $a_j$ is
present in query $q_i$, and to zero otherwise. A simple example of
query-attribute  matrix based on the workload excerpt from
Figure~\ref{exemple-charge} is featured in Table~\ref{exemple-matrice}.

\begin{figure}[hbt]
    \centering
        \begin{tabular}{ll}
            \hline
            $q_1$   & \small{\texttt{SELECT $F$.$a_1$, SUM($F$.$a_2$) FROM $F$, $D_1$}}\\
                        & \small{\texttt{WHERE $F$.$a_1$ = $D_1$.$a_3$ AND $D_1$.$a_4$ < 2000}}\\
                        & \small{\texttt{GROUP BY $F$.$a_1$}}\\
            $q_2$   & \small{\texttt{SELECT $F$.$a_1$, $F$.$a_5$, AVG($F$.$a_6$) FROM $F$, $D_1$, $D_2$}}\\
                        & \small{\texttt{WHERE $F$.$a_1$ = $D_1$.$a_3$ AND $F$.$a_5$ = $D_2$.$a_7$ AND $D_2$.$a_8$ = 'ABC'}}\\
                        & \small{\texttt{GROUP BY $F$.$a_1$, $F$.$a_5$}}\\
            $q_3$   & \small{\texttt{SELECT $F$.$a_1$, $F$.$a_9$, SUM($F$.$a_2$) FROM $F$, $D_1$, $D_3$}}\\
                        & \small{\texttt{WHERE $F$.$a_1$ = $D_1$.$a_3$ AND $F$.$a_9$ = $D_3$.$a_{10}$}}\\
                        & \small{\texttt{GROUP BY $F$.$a_1$, $F$.$a_9$}}\\
            \hline
        \end{tabular}
    \caption{Workload excerpt}
    \label{exemple-charge}
\end{figure}

\begin{table}[hbt]
    \centering
        \begin{tabular}{c|c|c|c|c|c|c|c|c|}
        \cline{2-9}
             & $a_1$ & $a_3$ & $a_4$ & $a_5$ & $a_7$ & $a_8$ & $a_9$ & $a_{10}$ \\
        \hline
        \multicolumn{1}{|c|}{$q_1$} & 1 & 1 & 1 & 0 & 0 & 0 & 0 & 0 \\
        \hline
        \multicolumn{1}{|c|}{$q_2$} & 1 & 1 & 0 & 1 & 1 & 1 & 0 & 0 \\
        \hline
        \multicolumn{1}{|c|}{$q_3$} & 1 & 1 & 0 & 0 & 0 & 0 & 1 & 1 \\
        \hline
        \end{tabular}
    \caption{Sample query-attribute matrix}
    \label{exemple-matrice}
\end{table}

This data structure directly corresponds to attribute-value tables that are
exploited by data mining algorithms. Here, attributes are queries and values,
attributes. Applying a data mining technique onto the query-attribute matrix
helps obtain a set of candidate objects (materialized views and indexes) $O_c$.

\subsection{Final object configuration construction}
\label{sec:FinalObjectConfiguration}

Our final materialized view and index configuration construction algorithm
(Figure~\ref{glouton}) is based on an ascending greedy search within the input
candidate object set $O_C$. It starts from an empty final object configuration
$O$, and then adds in it object $o$ from $O_C$ that maximizes objective
function $f_O$, at each iteration. For each object $o \in O_C$, the value of
$f_O(o)$ depends on objects already selected in $O$. Thus, it must be
recomputed at each iteration, which helps take view-index interactions into
account. The algorithm ends when objective function $f_O$ cannot be improved
any more, when there are no more candidate objects in $O_C$, or when storage
space $S$ allocated by the warehouse administrator to materialized views and
index is full.

\begin{figure}[hbt]
    \centering
    \begin{tabular}{|p{10cm}|}
    \hline
    \small{{\tt
    $O = \emptyset$ \newline
    $s_O = 0$ \newline
    \underline{Repeat} \newline
        \hspace*{0.5cm} $o_{max} = \emptyset$ \newline
        \hspace*{0.5cm} $f_{max} = 0$ \newline
        \hspace*{0.5cm} \underline{For each} $o \in O_C$ \underline{do} \newline
            \hspace*{1cm} \underline{If} $f_O(o) > f_{max}$ \underline{then} \newline
                \hspace*{1.5cm} $o_{max} = o$ \newline
                \hspace*{1.5cm} $f_{max} = f_O(o)$ \newline
            \hspace*{1cm} \underline{End if} \newline
        \hspace*{0.5cm} \underline{End for} \newline
        \hspace*{0.5cm} \underline{If} $f_O(o_{max}) > 0$ \underline{then} \newline
            \hspace*{1cm} $O = O \cup \{o_{max}\}$ \newline
            \hspace*{1cm} $s_O = s_O + size(o_{max})$ \newline
            \hspace*{1cm} $O_C = O_C \setminus \{o_{max}\}$ \newline
        \hspace*{0.5cm} \underline{End if} \newline
    \underline{Until} $f_O(o_{max}) \leq 0$ \underline{or} $O_C  = \emptyset$ \underline{or} $s_O \geq S$
    }}\\
    \hline
    \end{tabular}
    \caption{Final object configuration construction algorithm}
    \label{glouton}
\end{figure}

For a given workload $Q$ and an object configuration $O$, the objective
function $f_O$ may generally be expressed as follows: $f_O(o) = \alpha_o
~benefit_O(o) - \beta_o ~maintenance(o)$. Generally, $benefit_O(o) =
\frac{cost(Q, O) - cost(Q, O \cup \{o\})}{size(o)}$. However, taking view-index
interactions into account complicates this function's computation
(Section~\ref{sec:CostModels}).

Cost models developed in Section~\ref{sec:Applications} help compute the $cost$
and $maintenance$ functions. Coefficient $\alpha_o$ helps ponder benefit. It is
generally equal to one, but may also help favor index that avoid join
operations (Section~\ref{sec:FrequentItemsetMiningBasedIndexSelection}).
Finally, coefficient $\beta_o = |Q| p(o)$ is an estimator for the number of
updates of object $o$. The update probability of object $o$, $p(o)$, is equal
to $\frac{1}{|O|} \frac{\%refreshment}{\%interrogation}$, where
$\frac{\%refreshment}{\%interrogation}$ represents the proportion of warehouse
updates with respect to interrogations.

\section{Applications}
\label{sec:Applications}

This section presents three instances of our automatic data warehouse
performance optimization approach: automatic materialized view selection,
automatic index selection, and automatic, joint materialized view and index
selection. We particularly detail this last, newest application. Moreover, we
particularly insist, for each application, on its specificities in terms of
candidate object selection (e.g., the data mining technique we exploited) and cost
models used in building the final object configuration.

\subsection{Clustering-based materialized view selection}
\label{sec:ClusteringBasedMaterializedViewSelection}

In this application, we propose to select materialized views by clustering
queries from workload $Q$. Several syntactically similar queries have indeed a
high probability of being resolved by one single materialized view. Then, we
must build classes of similar queries from $Q$. Since the number of classes is
\emph{a priori} unknown, we have selected an unsupervised clustering
method.

Our approach's principle is similar to SQL workload compression
\cite{chaudhuri02}, a technique proposed in the relational database context to
optimize, for instance, index selection or approximate answer to aggregation
queries. We adapted this idea to the context of relational (with an SQL
decision-support workload) and XML (with an XQuery decision-support workload)
data warehouses \cite{adbis06,csit06}.

The main improvement brought by our approach lies at the candidate view
selection level. Most anterior methods indeed build a lattice or graph of all
syntactically correct views for a given workload. However, in practice, such
data structures are complex to build and browse. Using a clustering algorithm
helps drastically reduce the number of candidate materialized views by
proposing only a couple of views per class (only one in the best case ---
Section~\ref{selvues}) instead of one view per workload query. This
dimensionality reduction helps improve the whole process' efficiency  and
offers true scaling up capability.

\subsubsection{Candidate materialized view selection}
\label{selvues}

\paragraph{Query similarity and dissimilarity.} To perform clustering and check out
whether query classes are homogeneous, we must define query similarity and
dissimilarity measures. Let $M$ be a query-attribute matrix of general term
$m_{ij}$, defined on query set $Q = \{q_i, i=1..n\}$ and attribute set $A =
\{a_j, j = 1..l\}$. We define the elementary similarly and dissimilarity
between two queries $q_i$ and $q_{i'}$, regarding attribute $a_j$, as follows.

$\partial_{sim/a_j}(q_i, q_{i'}) = \left \{
\begin{array}{l}
1 $~if~$ m_{ij} = m_{i'j} = 1\\
0 $~otherwise$
\end{array}
\right.$
~~~~~
$\partial_{dissim/a_j}(q_i, q_{i'}) = \left \{
\begin{array}{l}
1 $~if~$ m_{ij} \neq m_{i'j}\\
0 $~otherwise$
\end{array}
\right.$

Note that these definitions are not symmetric. The absence of a given attribute
in two queries does indeed not constitute an element of similarity, unlike its
presence. We now extend these definitions onto attribute set $A$ to obtain
global similarity and dissimilarity between queries $q_i$ and $q_{i'}$.

$sim(q_i, q_{i'}) = \displaystyle{\sum_{j = 1}^l \partial_{sim/a_j}(q_i, q_{i'})}$
~~~~~
$dissim(q_i, q_{i'}) = \displaystyle{\sum_{j = 1}^l \partial_{dissim/a_j}(q_i, q_{i'})}$

\paragraph{Query clustering.} The objective of clustering is to build a natural
partition of queries that reflects their internal structure. Objects in the same
class must be strongly similar, while objects from different classes must be
strongly dissimilar. Let $P = \{C_k, k=1..p\}$ be a partition of $p$ classes (query sets).
We define interclass similarity between two distinct classes $C_a$ and $C_b$ from $P$, as well
as intraclass dissimilarity within class $C_a$ from $P$, as follows.

$Sim(C_a, C_b) = \displaystyle{\sum_{q_i \in C_a, q_{i'} \in C_b, } sim(q_i, q_{i'})}$
~~~~~
$Dissim(C_a) = \displaystyle{\sum_{q_i \in C_a, q_{i'} \in C_a, i < i'} dissim(q_i, q_{i'})}$

Eventually, we define on $P$ a measure of clustering quality $Q(P)$ that
helps capture the partition's natural aspect. $Q(P)$ indeed possesses low values for partitions
that have a strong intraclass homogeneity and a strong interclass disparity. $Q(P)$ must be minimized.

$Q(P) = \displaystyle{\sum_{a = 1..p, b = 1..p, a < b} Sim(C_a, C_b)} + \displaystyle{\sum_{a=1}^z Dissim(C_a)}$

To actually perform clustering, we selected the Kerouac algorithm \cite{jou03ker} that
bears interesting features in our context. It can indeed easily take our quality measure $Q(P)$ into account,
as well as integrate constraints in the clustering process. It is thus possible to satisfy a
precondition in the materialized view fusion process (see next paragraph): queries from
one given class must share the same joining conditions.

\paragraph{Candidate view fusion.} The output of clustering is a set of similar query classes.
Our objective is to associate to each class the smallest possible number of
materialized views that cover all the class' queries. To achieve this goal, we
consider each query as a potential view and run a fusion process to decrease
their number. The algorithm we use is very similar to the one proposed by
\citeA{agr00aut}. However, in our context, it is much more efficient since it
is applied onto a limited number of views in each class instead of the whole
set of candidate views derived from the workload. The output of fusion applied
on classes obtained in the previous step is the set of candidate materialized
views.

\subsubsection{Cost models}
\label{CMV}

In most of the (relational) data warehouse cost models from the literature, the
cost of a query $q$ is supposed proportional to the size (in tuples) of the
materialized view exploited by $q$~\cite{gol98met}. The same assumption is made
for view maintenance cost. Hence, we reuse a model that estimates the size of a
given materialized view. It has been proposed by \citeA{gol98met} and exploits
\citeA{yao77app}'s formula to estimate the number of tuples $|V|$ of a view $V$
composed of $k$ attributes $a_1, a_2, ..., a_k$ and based on fact table $F$ and
$d$ dimensions $D_1, D_2, ..., D_d$: $|V| = max\_size(V) \times \left ( 1 -
\displaystyle{\prod_{i = 1}^{|F|}\frac{max\_size(F) \times (1 -
\frac{1}{max\_size(V)}) - i + 1}{max\_size(F) - i + 1}} \right )$, where
$max\_size(V) = \displaystyle{\prod_{i = 1}^k |a_i|}$ and $max\_size(F) =
\displaystyle{\prod_{i = 1}^d |D_i|}$.

When ratio $\frac{max\_size(F)}{max\_size(V)}$ is high enough, \citeA{car75ana}'s formula helps obtain a good approximation:
$|V| = max\_size(V) \times \left ( 1 - \left (1 - \frac{1}{max\_size(V)} \right )^{|F|} \right )$.

$V$'s size in bytes is then $size(V) = |V| \times \displaystyle{\sum_{i = 1}^n size(d_i)}$, where $size(d_i)$ is the
size in bytes of dimension $d_i$ from $V$ (which can be directly obtained from the warehouse metadata) and $n$ the
number of dimensions in $V$.  Yao and Cardenas' formulae assume data are uniformly
distributed and tend to overestimate view size. However, they are easy to implement and
fast to compute. Other, more precise methods exploit data sampling and statistical
laws \cite{shu96sto,cha99sam,nad01par}, but they are much harder to implement.

Eventually, this cost model is very easy to adapt to the XML context by
establishing equivalences between relations and XML documents on one hand, and
tuples and XML elements on the other hand. The only true difference lies in
$size(d_i)$'s computation, but it is also obtained from warehouse metadata in
the XML context.

\subsection{Frequent itemset mining-based index selection}
\label{sec:FrequentItemsetMiningBasedIndexSelection}

In this application, we work on optimizing the execution of join operations in a decision-support query workload.
We propose an index selection based on the extraction from the workload of frequent attributes that may support indexes.

We have first worked on classical, B-tree-like indexes \cite{aou03fre}. We
focus in this paper on bitmap index selection \cite{aou05aut}. These data
structures \cite{nei95mul} are particularly adapted to the data warehouse
context. They indeed render logical and counting operations efficient (they
operate directly on bitmaps stored in the main memory), and help precompute
join operations at index creation time. Moreover, bitmap storage space is
small, especially when the indexed attributes' cardinality is low, which is
usually the case in a warehouse's dimensions.

Our approach's originality mainly lies in the use of frequent itemset mining for
selecting the most pertinent candidate indexes. However, it also has another
advantage. The few approaches that help select multi-attribute indexes exploit
an iterative process to build them: mono-attribute indexes in the first
iteration, 2-attribute indexes in the second, and so on \cite{cha97eff}. In our
approach, frequent itemsets, which are attribute sets of variable size, help
directly propose multi-attribute candidate indexes. Furthermore, these
candidate indexes are \emph{a priori} pertinent, while combinations generated
from smaller candidate indexes are not necessarily all pertinent. Thus, our
approach avoids pruning them by providing a smaller set of pertinent
candidates.

Eventually, most existing index selection techniques (Section~\ref{soa-index})
only exploit B-tree indexes. Though this type of index is widely used in DBMSs,
it is not the best adapted to index voluminous data and low cardinality
attributes. In the data warehouse context, bitmap join indexes we privilege are
more efficient.

\subsubsection{Candidate index selection}
\label{CIS}

When building the extraction context (query-attribute matrix) that is exploited
by a data mining algorithm to select candidate indexes, we use knowledge
relative to database administration and performance optimization, much like
\citeA{fel03nea}. Such an attribute preselection helps reduce the mining
algorithm search space and, mechanically, improves its response time.

Knowledge is formalized under the form of ``if-then'' rules, e.g., ``if a
predicate is like $attribute \neq value$, then $attribute$ must not be
selected''. Such a predicate would indeed not exploit an index defined on
$attribute$, all its values being scanned but $value$.

We base the final selection of candidate indexes on the intuition that the
importance of an attribute to index is strongly correlated to its appearance
frequency in the workload. Frequent itemset mining \cite{agr94fast} appears as
a natural solution to extract these attributes. Many frequent itemset mining
algorithms are available in the literature. We selected Close \cite{pas99dis},
which presents several advantages in our context.

First, Close helps process voluminous workloads. It indeed exploits Galois
closure operators, which reduce the number of accesses to the extraction
context when searching for frequent itemsets. Close is also efficient when the
extraction context is dense, which is our case, since query sets often form
logical suites. Eventually, closed frequent itemsets extracted by this
algorithm are fewer than all frequent itemsets (which can nonetheless be
generated from the closed frequent itemsets). This helps reduce computing time
and avoid multiplying useless candidate indexes.

\subsubsection{Cost models}
\label{CMI}

\paragraph{Data access cost through a bitmap join index.} Data access is performed in two steps: scanning
the index' bitmaps, and then reading the tuples. If access to the bitmaps
is direct and data are uniformly distributed, which is a reasonable assumption
according to \citeA{cho93ind}, index traversal cost is $d \frac{|A| |F|}{8 S_p}$. $d$ is the number
of predicates applied on indexed attribute $A$.  $F$ is the fact table. $S_p$ is the size of a
disk page. $\frac{|A| |F|}{8}$ represents the size of the bitmap index \cite{wu98enc}.

The number of tuples read by a query using $d$ bitmaps is $d \frac{|F|}{|A|}$
if data are uniformly distributed. The number of corresponding input/output is
then equal to $p_F (1 - e^{-\frac{d |F|}{p_F |A|}})$ \cite{nei97imp}, where
$p_F$ is the number of disk pages that are necessary to store $F$. Finally,
$cost = d \frac{|A| |F|}{8 S_p} + p_F (1 - e^{-\frac{d |F|}{p_F |A|}})$.

If bitmap access is performed through a B-tree, as is the case in the Oracle
DBMS, for instance, B-tree descent cost must be taken into account: $log_m |A|
- 1$, where $m$ is the B-tree order. Leaf nodes traversal cost is then
$\frac{|A|}{m-1}$ at worst. However, bitmap index traversal cost is reduced to
$d \frac{|F|}{8 S_p}$. Then, $cost = log_m |A| - 1 + d \frac{|F|}{8 S_p} + p_F
(1 - e^{-\frac{d |F|}{p_F |A|}})$.

\paragraph{Bitmap join index maintenance cost.} Let a bitmap join index be defined on
attribute $A$ from dimension $D$. When inserting a tuple into fact table $F$,
$D$ must be traversed to find the tuple that must be joined to the one inserted
in $F$: $p_D$ pages are read. Then, the index' bitmaps must be updated. At
worst, they are all traversed and $\frac{|A| |F|}{8 S_p}$ pages are read.
Hence, $maintenance_F = p_D + \frac{|A| |F|}{8 S_p}$.

When inserting a tuple into dimension $D$, update may be without domain
expansion, then a bit corresponding to the inserted tuple must be added to each
bitmap; or with domain expansion, then a new bitmap must be created. Then,
$maintenance_D = p_F + (1 + \xi) \frac{|A| |F|}{8 S_p}$, where $\xi = 1$ when
expanding the domain and $\xi = 0$ otherwise.

\subsection{Joint materialized view and index selection}
\label{sec:JointMaterializedViewAndIndexSelection}

In this eventual application, we seek to select a configuration of materialized
views and indexes that are mutually beneficial, in order to further optimize
the response time of decision-support queries. More precisely, we aim at truly
taking view-index interactions into account and at optimizing storage space
sharing between materialized views and indexes. Existing approaches indeed
consider indexes and materialized views as distinct objects, whose benefit and
maintenance cost are invariant and independent from already-selected objects.
Moreover, few consider indexing materialized views.

\citeA{bel00eff}'s approach, which is closest to ours, starts from an initial
solution composed of indexes and materialized views isolately selected under
storage space constraint. Taking this constraint into account \emph{a priori}
might eliminate solutions that are susceptible to become pertinent in the next
iterations of the selection process. Hence, we only introduce the storage space
constraint \emph{a posteriori}, within the selection algorithm. Furthermore,
object replacement policies in storage spaces respectively allotted to indexes
and materialized views exploit these objects' usage frequency, and not the
benefit brought by their simultaneous usage.

\subsubsection{Candidate object selection}
\label{sec:CandidateObjectSelection}

\begin{figure}[hbt]
{\centering\resizebox*{0.75\textwidth}{!}{\includegraphics{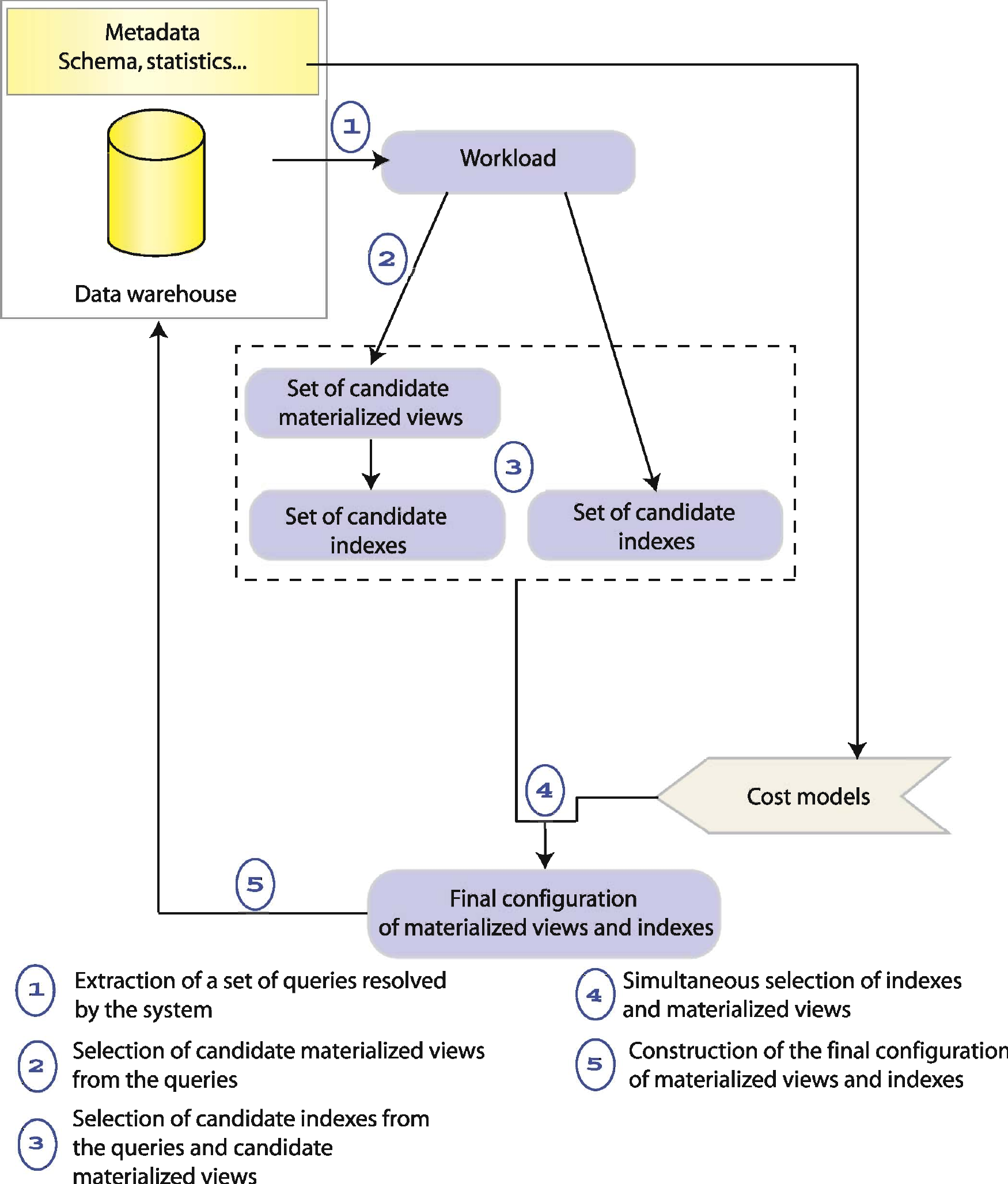}}
\par}
\caption{Joint materialized view and index selection approach}\label{fig:archi}
\end{figure}

First, let us detail and specialize the automatic performance optimization
strategy presented in Section~\ref{sec:GeneralPrinciple} for joint materialized
view and index selection. Here, we exploit the modular structure of our approach: our
input is a set of candidate objects (materialized views and indexes) obtained
with any existing selection algorithm, such as the ones we propose. Then, we
exploit specific data structures and cost models to recommend a pertinent
configuration of materialized views and indexes through the following steps
(Figure~\ref{fig:archi}):
\begin{enumerate}
\item extract a representative query set $Q$ from system workload;
\item build a set of candidate materialized views $V_C$ using the approach described in Section~\ref{selvues}, with $Q$ as input;
\item build a set of candidate indexes $I_C$ using the approach described in Section~\ref{CIS}, with $Q \cup V_C$ as input;
\item simultaneously select materialized views and indexes from $O_C = V_C \cup I_C$;
\item build the final configuration of materialized views and indexes $O \subseteq O_C$ under storage space constraint $S$.
\end{enumerate}

\subsubsection{Specific data structures}
\label{sec:DataStructures}

After building the set of candidate materialized views, indexes and indexes on
views $O_C$, we aim at combining them to recommend a pertinent configuration of
materialized views and indexes $O$. To consider the relationships between these
objects in this process, we need to materialize them. For this purpose, we use
three binary matrices: the ``query-view'' matrix, the ``query-index'' matrix
and the ``view-index'' matrix that we detail in the following paragraphs.

To better illustrate how these data structures are designed, let us consider
the workload sample from Figure~\ref{fig:charge_sim}. Candidate materialized
views and indexes obtained from this workload by applying our strategy are
featured in Figures~\ref{fig:vue_sim} and \ref{fig:index_sim}, respectively.

\begin{figure}[hbt]
\centering{ \tiny{
\begin{tabular}{clcl}\hline
$q_{1}$ & \textbf{select} sales.time\_id, \textbf{sum}(amount\_sold) & $q_{5}$ & \textbf{select} promotions.promo\_name, \\
        & \textbf{from} sales, times & & \textbf{sum}(amount\_sold) \\
        & \textbf{where} sales.time\_id = times.time\_id & &  \textbf{from} sales, promotions\\
        & \textbf{and} times.time\_fiscal\_year = 2000 & &  \textbf{where} sales.promo\_id = promotions.promo\_id\\
        & \textbf{group by}  sales.time\_id  & & \textbf{and} promotions.promo\_begin\_date=`30/01/2000' \\
        & & & \textbf{and} promotions.promo\_end\_date=`30/03/2000' \\
        & & & \textbf{group by} promotions.promo\_name \\ \\
$q_{2}$ & \textbf{select} sales.prod\_id,  & $q_{6}$ & \textbf{select} customers.cust\_marital\_status, \\
        & \textbf{sum}(amount\_sold) & &  \textbf{sum}(quantity\_sold)\\
        & \textbf{from} sales, products, promotions  & & \textbf{from} sales, customers, products\\
        & \textbf{where} sales.prod\_id = products.prod\_id & & \textbf{where} sales.cust\_id = customers.cust\_id \\
        & \textbf{and} sales.promo\_id = promotions.promo\_id & & \textbf{and} sales.prod\_id = products.prod\_id\\
        & \textbf{and} promotions.promo\_category = `news paper' & & \textbf{and} customers.cust\_gender = `woman' \\
        & \textbf{group by} sales.prod\_id & & \textbf{and} products.prod\_name = `shampooing'\\
        & & &  \textbf{group by} customers.cust\_first\_name \\ \\
$q_{3}$ & \textbf{select} customers.cust\_gender, \textbf{sum}(amount\_sold) & $q_{7}$ &  \textbf{select} products.prod\_name, \textbf{sum}(amount\_sold)\\
        & \textbf{from} sales, customers, products, & & \textbf{from} sales, products, promotions \\
        & \textbf{where} sales.cust\_id = customers.cust\_id & & \textbf{where} sales.prod\_id = products. prod\_id \\
        & \textbf{and} sales.prod\_id = products.prod\_id & & \textbf{and} sales.promo\_id =promotions.promo\_id \\
        & \textbf{and} customers.cust\_marital\_status =`single'& & \textbf{and} products.prod\_category=`tee shirt' \\
        & \textbf{and} products.prod\_category = `women' & & \textbf{and} promotions.promo\_end\_date=`30/04/2000' \\
        & \textbf{group by} customers.cust\_gender & & \textbf{group by} products.prod\_name
        \\\\
$q_{4}$ & \textbf{select} products.prod\_name, sum(amount\_sold) & $q_{8}$ & \textbf{select} channels.channel\_desc, sum(quantity\_sold)\\
        & \textbf{from} sales, products, promotions & & \textbf{from} sales, channels\\
        & \textbf{where} sales.prod\_id = products.prod\_id & & \textbf{where} sales.channel\_id = channels.channel\_id \\
        & \textbf{and} sales.promo\_id = promotions.prom\_id & & \textbf{and} channels.channel\_class = `Internet' \\
        & \textbf{and} promotions.promo\_category = `TV' & & \textbf{group by} channels.channel\_desc\\
        & \textbf{group by} products.prod\_name & & \\
\hline
\end{tabular}
}} \caption{Sample workload}\label{fig:charge_sim}
\end{figure}

\begin{figure}[hbt]
\centering{ \tiny{
\begin{tabular}{cl}\hline
$v_{1}$ & \textbf{create} \textbf{materialized view} $v_{1}$ \textbf{as} \\
        & \textbf{select} sales.time\_id, times.time\_fiscal\_year, \\
        & \textbf{sum}(amount\_sold)  \\
        & \textbf{from} sales, times \\
        & \textbf{where} sales.time\_id = times.time\_id \\
        & \textbf{group by}  sales.time\_id, times.times\_fiscal\_year  \\ \\
$v_{2}$ & \textbf{create} \textbf{materialized view} $v_{2}$ \textbf{as} \\
        & \textbf{select} sales.prod\_id, sales.cust\_id, channels.channel\_desc, \\
        & channels.channel\_class, \textbf{sum}(quantity\_sold) \\
        & \textbf{from} sales, channels, products, customers \\
        & \textbf{where} sales.prod\_id = products.prod\_id \\
        & \textbf{and} sales.channel\_id = channels.channel\_id \\
        & \textbf{and} sales.cust\_id = customers.cust\_id \\
        & \textbf{group by} sales.prod\_id, sales.cust\_id, channels.channel\_desc, \\
        & channels.channel\_class \\ \\
$v_{3}$ &\textbf{create} \textbf{materialized view} $v_{3}$ \textbf{as}\\
        & \textbf{select} customers.cust\_first\_name, products.prod\_name,  \\
        & products.prod\_category, customers.cust\_gender,  \\
        & customers.cust\_marital\_status, \textbf{sum}(sales.quantity\_sold)  \\
        & \textbf{from} sales, customers, products  \\
        & \textbf{where} sales.cust\_id = customers.cust\_id \\
        & \textbf{and} sales.prod\_id = products.prod\_id \\
        & \textbf{group by} customers.cust\_first\_name, products.prod\_name,  \\
        & products.prod\_category, customers.cust\_gender, \\
        & customers.cust\_marital\_status  \\ \\
$v_{4}$ &\textbf{create} \textbf{materialized view} $v_{4}$ \textbf{as}  \\
        & \textbf{select} products.prod\_name, products.prod\_category,   \\
        & promotions.promo\_category, \textbf{sum}(amount\_sold)   \\
        & \textbf{from} sales, products, promotions  \\
        & \textbf{where} sales.prod\_id = products.prod\_id    \\
        & \textbf{and} sales.promo\_id = promotions.promo\_id   \\
        & \textbf{group by} products.prod\_name, products.prod\_category,  \\
        & promotions.promo\_category   \\ \\
 $v_{5}$ &\textbf{create} \textbf{materialized view} $v_{5}$ \textbf{as}\\
          &   \textbf{select} sales.prod\_id, products.prod\_category,\\
         &  promotions.promo\_category, \textbf{sum}(amount\_sold) \\
         &  \textbf{from} sales, products, promotions\\
         &  \textbf{where} sales.prod\_id = = products.prod\_id\\
         & \textbf{and} sales.promo\_id = promotions.promo\_id\\ \\
$v_{6}$ & \textbf{create} \textbf{materialized view} $v_{6}$ \textbf{as}\\
        &  \textbf{select} channels.channel\_class, products.prod\_name, channels.channel\_desc,\\
        & products.prod\_category, \textbf{sum}(sales.quantity\_sold), \textbf{sum}(sales.amount\_sold)\\
        & \textbf{from} sales, channels, products \\
        & \textbf{where} sales.prod\_id = products.prod\_id \\
        & \textbf{and} sales.channel\_id = channels.channel\_id\\
        &  \textbf{group by} channels.channel\_class, products.prod\_name,\\
        & products.prod\_category, channels.channel\_desc\\ \\
$v_{7}$ &  \textbf{create} \textbf{materialized view} $v_{7}$ \textbf{as}\\
        & \textbf{select} sales.prod\_id, products.prod\_category, \\
        & channels.channel\_desc, promotions.promo\_name,  \\
        & promotions.promo\_begin\_date, promotions.promo\_end\_date, \\
        &  products.prod\_name, \textbf{sum}(sales.quantity\_sold), \textbf{sum}(sales.amount\_sold) \\
        &  \textbf{from} sales, products, promotions\\
        &  \textbf{where} sales.prod\_id = products.prod\_id \\
        &  \textbf{and} sales.promo\_id = promotions.promo\_id \\
        & \textbf{and} sales.channel\_id = channels.channel\_id\\
        &  \textbf{group by} sales.prod\_id, products.prod\_category, channels.channel\_desc, \\
        & promotions.promo\_name, promotions.promo\_begin\_date, \\
        & promotions.promo\_end\_date, products.prod\_name\\ \\
 \hline
\end{tabular}
}} \caption{Candidate materialized views}\label{fig:vue_sim}
\end{figure}

\begin{figure}[hbt]
\centering{\small{
\begin{tabular}{cl}\hline
\textbf{Indexes} & \textbf{Indexed attributes}\\ \hline
$i_{1}$ & promotions.promo\_category \\
$i_{2}$ & channels.channel\_desc \\
$i_{3}$ & channels.channel\_class \\
$i_{4}$ & customers.cust\_marital\_status \\
$i_{5}$ & customers.cust\_gender \\
$i_{6}$ & times.time\_begin\_date \\
$i_{7}$ & times.time\_end\_date \\
$i_{8}$ & times.fiscal\_year \\
$i_{9}$ & products.prod\_name \\
$i_{10}$ & products.prod\_category \\
$i_{11}$ & promotions.promo\_name \\
$i_{12}$ & customers.cust\_first\_name \\
\hline
\end{tabular}
}}\caption{Candidate indexes}\label{fig:index_sim}
\end{figure}

\paragraph{Query-view matrix.}
The query-view matrix ($QV$) captures existing relationships between workload
queries and the materialized views extracted from these queries, i.e., views
that are exploited by at least one workload query. This matrix may be viewed as
the result of rewriting queries with respect to candidate materialized views.
The query-view matrix' rows and columns are workload queries and candidate
materialized views, respectively. Its general term $QV_{qv}$ is equal to one if
a given query $q$ exploits the corresponding view $v$, and to zero otherwise.
Table~\ref{tab:req_vue} presents the query-view matrix corresponding to the
example from Figures~\ref{fig:charge_sim} and \ref{fig:vue_sim}.

\begin{table}[hbt]
\centering
\begin{tabular}{|c|c|c|c|c|c|c|c|}
\cline{2-8}
\multicolumn{1}{c|}{}  & $v_{1}$ & $v_{2}$ & $v_{3}$ &  $v_{4}$ & $v_{5}$ &
$v_{6}$ & $v_{7}$ \\ \hline
$q_{1}$ & 1 & 0 & 0 & 0 & 0 & 0 & 0  \\ \hline
$q_{2}$ & 0 & 0 & 0 & 1 & 0 & 0 & 0  \\ \hline
$q_{3}$ & 0 & 0 & 1 & 0 & 0 & 0 & 0  \\ \hline
$q_{4}$ & 0 & 0 & 0 & 1 & 0 & 0 & 0  \\ \hline
$q_{5}$ & 0 & 0 & 0 & 0 & 0 & 0 & 1  \\ \hline
$q_{6}$ & 0 & 0 & 1 & 0 & 0 & 0 & 0  \\ \hline
$q_{7}$ & 0 & 0 & 0 & 0 & 0 & 0 & 1  \\ \hline
$q_{8}$ & 0 & 1 & 0 & 0 & 0 & 1 & 0  \\ \hline
\end{tabular}
\caption{Sample query-view matrix}\label{tab:req_vue}
\end{table}

\paragraph{Query-index matrix.}
The query-index matrix ($QI$) stores the indexes built on base tables. This
matrix may be viewed as the result of rewriting queries with respect to
candidate indexes. The query-index matrix' rows and columns are workload
queries and candidate indexes, respectively. Its general term $QI_{qi}$ is
equal to one if a given query $q$ exploits the corresponding index $i$, and to
zero otherwise. Table~\ref{tab:req_index} presents the query-index matrix
corresponding to the example from Figures~\ref{fig:charge_sim} and
\ref{fig:index_sim}.

\begin{table}[hbt]
\centering
\begin{tabular}{|c|c|c|c|c|c|c|c|c|c|c|c|c|}
\cline{2-13}
\multicolumn{1}{c|}{}  & $i_{1}$ & $i_{2}$ & $i_{3}$ &  $i_{4}$ & $i_{5}$ &
$i_{6}$ & $i_{7}$ & $i_{8}$ & $i_{9}$ & $i_{10}$ &
      $i_{11}$  & $i_{12}$ \\ \hline
$q_{1}$ & 0 & 0 & 0 & 0 & 0 & 0 & 0 & 1 & 0 & 0 & 0 & 0 \\ \hline
$q_{2}$ & 1 & 0 & 0 & 0 & 0 & 0 & 0 & 0 & 0 & 0 & 0 & 0\\ \hline
$q_{3}$ & 0 & 0 & 0 & 1 & 1 & 0 & 0 & 0 & 0 & 1 & 0 & 0 \\ \hline
$q_{4}$ & 1 & 0 & 0 & 0 & 0 & 0 & 0 & 0 & 1 & 0 & 0 & 0 \\ \hline
$q_{5}$ & 0 & 0 & 0 & 0 & 1 & 1 & 0 & 0 & 0 & 0 & 1 & 0 \\ \hline
$q_{6}$ & 0 & 0 & 0 & 0 & 1 & 0 & 0 & 0 & 1 & 0 & 0 & 1 \\ \hline
$q_{7}$ & 0 & 0 & 0 & 0 & 0 & 0 & 1 & 0 & 1 & 1 & 0 & 0 \\ \hline
$q_{8}$ & 0 & 1 & 1 & 0 & 0 & 0 & 0 & 0 & 0 & 0 & 0 & 0 \\ \hline
\end{tabular}
\caption{Sample query-index matrix}\label{tab:req_index}
\end{table}

\paragraph{View-index matrix.}
The view-index matrix ($VI$) identifies candidate indexes that are recommended
for candidate materialized views returned by our view selection algorithm. The
query-index matrix' rows and columns are candidate views and candidate indexes
on these views, respectively. Its general term $VI_{vi}$ is equal to one if a
given materialized view $v$ exploits the corresponding index $i$, and to zero
otherwise. Table~\ref{tab:vue_index} presents the view-index matrix
corresponding to the example from Figures~\ref{fig:vue_sim} and
\ref{fig:index_sim}.

\begin{table}[hbt]
\centering
\begin{tabular}{|c|c|c|c|c|c|c|c|c|c|c|c|c|}
\cline{2-13}
\multicolumn{1}{c|}{}  & $i_{1}$ & $i_{2}$ & $i_{3}$ &  $i_{4}$ & $i_{5}$ &
$i_{6}$ & $i_{7}$ & $i_{8}$ &  $i_{9}$ & $i_{10}$ & $i_{11}$ & $i_{12}$\\
\hline
$v_{1}$ & 0 & 0 & 0 & 0 & 0 & 0 & 0 & 1 & 0 & 0 & 0 & 0 \\ \hline
$v_{2}$ & 0 & 1 & 0 & 0 & 0 & 0 & 0 & 0 & 0 & 0 & 0 & 0 \\ \hline
$v_{3}$ & 0 & 0 & 0 & 1 & 1 & 0 & 0 & 0 & 1 & 1 & 0 & 1 \\ \hline
$v_{4}$ & 1 & 0 & 0 & 0 & 0 & 0 & 0 & 0 & 1 & 1 & 0 & 0 \\ \hline
$v_{5}$ & 1 & 0 & 0 & 0 & 0 & 0 & 0 & 0 & 0 & 1 & 0 & 0 \\ \hline
$v_{6}$ & 0 & 1 & 1 & 0 & 0 & 0 & 0 & 0 & 1 & 1 & 0 & 0  \\ \hline
$v_{7}$ & 0 & 1 & 0 & 0 & 0 & 1 & 1 & 0 & 1 & 1 & 1 & 0 \\ \hline
\end{tabular}
\caption{Sample view-index matrix}\label{tab:vue_index}
\end{table}

\subsubsection{Cost models}
\label{sec:CostModels}

We have already presented in Sections~\ref{CMV} and \ref{CMI} cost models
relative to materialized views and bitmap join indexes, respectively. Since
indexes defined on materialized views are generally B-trees or derivatives, we
first recall here the cost models that relate to these indexes. Then, we
discuss the benefit of view materialization vs. indexing.

\paragraph{Data access cost through a B-tree index.} Data access through an
index is subdivided into to steps: index traversal to find key values
corresponding to the query ($C_{traversal}$ cost), and then searching for these
identifiers in the database ($C_{search}$ cost). Let $q$ be a query, $\zeta$ a
set of indexes, $SNA_q$ the set of attributes that are present in query $q$'s
restriction clause (the \texttt{Where} clause in SQL), $BF_a$ the bloc factor
of the index built on attribute $a$ (the average number of $(key, identifier)$
couples per disk page), $SF_a$ the selectivity factor of attribute $a$, and
finally $v$ the accessed materialized view. Then: $C_{traversal} =
\displaystyle{\sum_{a \in (\zeta \cap SNA_q)} \lceil log_{BF_a}|v|\rceil +
\left \lceil \frac{SF_a |v|}{BF_a} \right \rceil - 1}$.

The number of identifiers to search for is then $N = |v| \displaystyle{\prod_{a
\in (\zeta \cap SNA_q)} SF_a}$. According, to \citeA{car75ana}'s formula, the
number of disk pages to access is: $C_{search} = S_p \left (1 - (1 -
\frac{1}{S_p})^N \right )$, where $S_p$ represents disk page size.

Finally, data access cost through a B-tree index is $cost = C_{traversal} + C_{search}$.

\paragraph{B-tree index maintenance cost.} Classically, this cost is expressed as follows:\\
$maintenance = \displaystyle{ \sum_{op \in \{ins, del, upd\}} \displaystyle{
\sum f_{op} \displaystyle{ \sum_{a \in A_{op}} C_{op}(a)} } }$; where
$f_{ins}$, $f_{del}$ and $f_{upd}$ are insert, delete and update frequencies,
respectively; and $C_{ins}(a)$, $C_{del}(a)$ and $C_{upd}(a)$ are maintenance
costs related to an insert, delete or update operation on attribute $a$,
respectively. $A_{op}$ is the set of considered attributes. $A_{ins} = A_{del}
= \zeta$, where $\zeta$ is the set of indexes to maintain. \\$A_{upd} = \zeta
\cap SNA_{upd}$, where $SNA_{upd}$ is the set of attributes to update. Finally,
maintenance costs are the following~\cite{whang85}: $C_{ins}(a) = C_{del}(a) =
\lceil log_{BF_a}|v|\rceil$
and\\
$C_{upd}(a) = \lceil log_{BF_a}|v|\rceil + \left \lceil \frac{|v| SF_a}{2 BF_a} \right \rceil - 1$.

\paragraph{View materialization and indexing benefit.} In the general
case (Section~\ref{sec:FinalObjectConfiguration}), the benefit brought by
selecting an object $o$ is defined as the difference between the execution cost
of query workload $Q$ before and after inserting $o$ into the final object
configuration $O$. Taking view-index relationships into account implies
redefining the benefit function. Let $i \in O_C$ and $v \in O_C$ be a candidate
index and a candidate materialized view, respectively. Adding $i$ or $v$ into
$O$ may lead to the benefit cases enumerated in Tables~\ref{benefindex}
and~\ref{benefvue}, respectively, depending on interactions between $i$ and
$v$.

\begin{table}[htb]
\centering{
\begin{tabular}{c|c|c|}
    \cline{2-3}
    & $VI_{vi} = 1$ & $VI_{vi} = 0$ \\
    \hline
    \multicolumn{1}{|c|}{$v \in O$} & $min$(materialization, indexing $v$) & indexing \\
    \hline
    \multicolumn{1}{|c|}{$v \not\in O$} & --- & indexing \\
    \hline
\end{tabular}}
\caption{Benefit brought by index $i$}
\label{benefindex}
\end{table}

\begin{table}[htb]
\centering{
\begin{tabular}{c|c|c|}
    \cline{2-3}
    & $VI_{vi} = 1$ & $VI_{vi} = 0$ \\
    \hline
    \multicolumn{1}{|c|}{$i \in O$} & --- & materialization \\
    \hline
    \multicolumn{1}{|c|}{$i \not\in O$} & $min$(indexing $v$, materialization) & materialization \\
    \hline
\end{tabular}}
\caption{Benefit brought by materialized view $v$}
\label{benefvue}
\end{table}

Indexing and view materialization benefits for $Q$, brought by adding index $i$
or view $v$ into $O$, respectively, may hence be expressed as follows.

$benefit_O(i) = \left \{
\begin{array}{l}
\frac{cost(Q, O) - cost(Q, O \cup \{i\})}{size(i)} $~if~$ VI_{vi} = 0 ~\forall ~v \in V ~(V \subseteq O)\\
\frac{cost(Q, O) - cost(Q, O \cup \{i\} \cup V')}{size(i) + \sum{v' \in V'} size(v')} $~if~$ V' = \{v \in V, VI_{vi} = 1\} \neq \emptyset\\
0 $~otherwise$
\end{array}
\right.$

$benefit_O(v) = \left \{
\begin{array}{l}
\frac{cost(Q, O) - cost(Q, O \cup \{v\})}{size(v)} $~if~$ VI_{vi} = 0 ~\forall ~i \in I ~(I \subseteq O)\\
\frac{cost(Q, O) - cost(Q, O \cup \{v\} \cup I')}{size(v) + \sum{i' \in I'} size(i')} $~if~$ I' = \{i \in I, VI_{vi} = 1\} \neq \emptyset\\
0 $~otherwise$
\end{array}
\right.$

\section{Experiments}
\label{sec:Experiments}

In order to experimentally validate our generic approach for optimizing data
warehouse performance, we have run several series of tests. We summarize the
main results in the following sections. Regarding isolate materialized view or
index selection, the interested reader can refer to
\cite{aou03fre,aou05aut,adbis06,csit06} for more complete results.

\subsection{Experimental conditions}
\label{sec:ExperimentalConditions}

All our tests have been run on a 1~GB data warehouse implemented within Oracle
$9i$, on a Pentium 2.4~GHz PC with 512~MB RAM and a 120~GB IDE disk. Our test
data warehouse is actually derived from Oracle's sample data warehouse. Its
star schema is composed of one fact table: \texttt{Sales}; and five dimensions:
\texttt{Customers}, \texttt{Products}, \texttt{Times}, \texttt{Promotions} and
\texttt{Channels}. The workload we executed on this data warehouse is composed
of 61 decision-support queries involving aggregation operations and multiple
joins between the fact table and dimension tables. Due to space constraints, we
do not reproduce here the full data warehouse schema nor the detail of each
workload query, but they are both available
on-line\footnote{\url{http://eric.univ-lyon2.fr/~kaouiche/adbis.pdf}}.

Note that our experiments are based on an ad-hoc benchmark because, at the time
we performed them, there was no standard benchmark for data warehouses.
TPC-H~\cite{tpch} does indeed not feature a true multidimensional schema and
thus does not qualify, and TPC-DS'~\cite{tpcds} draft specifications had not
been issued yet.

\subsection{Materialized view selection results}
\label{sec:MaterializedViewSelection}

\begin{figure}[hbt]
    \centering
    \includegraphics[width=10cm]{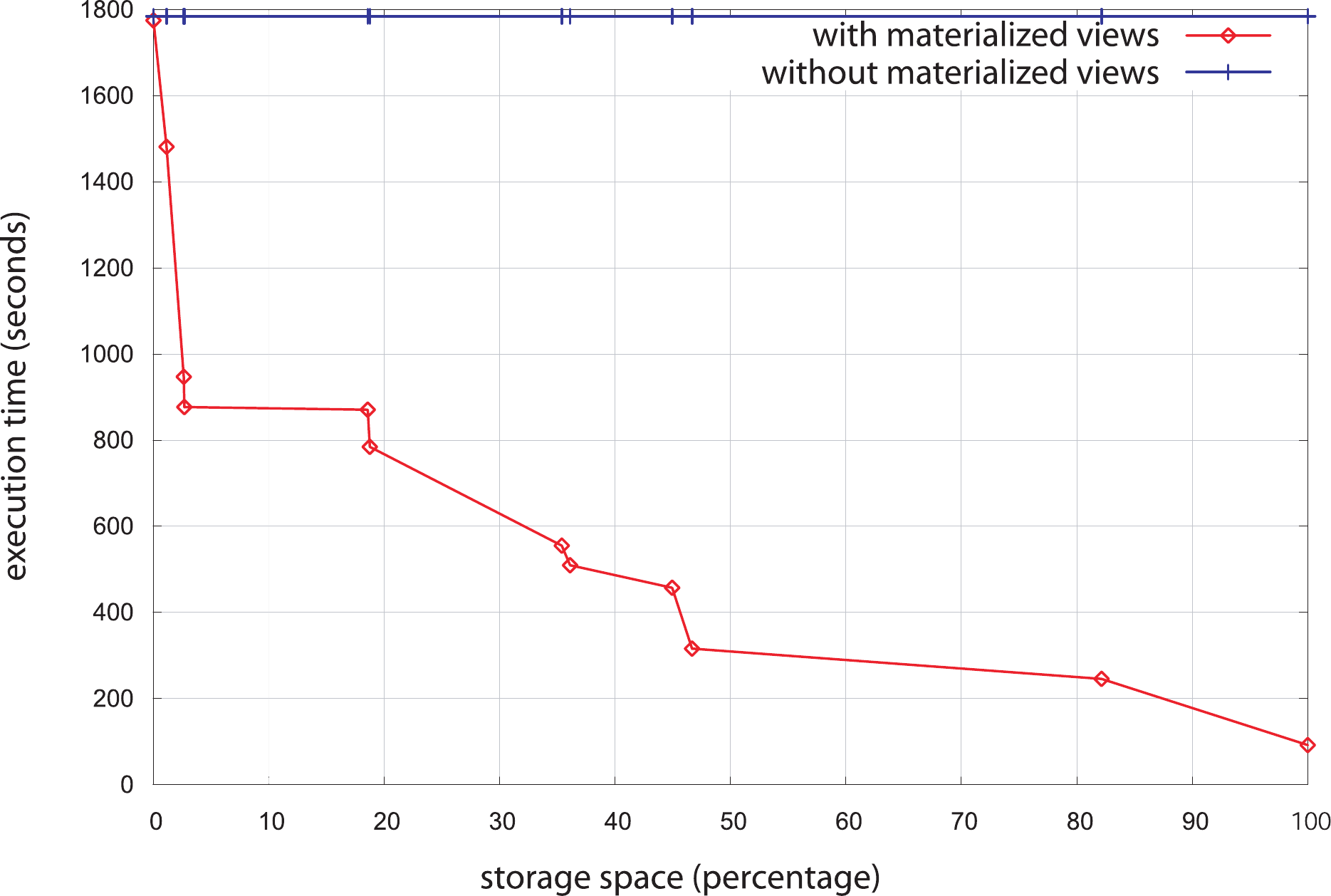}
    \caption{Views materialization experiment results}
    \label{fig:exp-mv}
\end{figure}

We plotted in Figure~\ref{fig:exp-mv} the variation of workload execution time
with respect to the storage space allotted for materialized views. This figure
shows that the views we select significantly improve query execution time.
Moreover, execution time decreases when storage space occupation increases.
This is predictable because we create more materialized views when storage
space $S$ is large and thereby better improve execution time. Let $S_V$ be the
disk space that is necessary to store all the candidate materialized views. The
average gain in performance is indeed 68.9\% when $S = 35.4\% \times S_V$. It
is equal to 94.9\% when $S = 100\% \times S_V$ (when the storage space
constraint is relaxed).

Moreover, we have demonstrated the relevance of the materialized views that are
selected with our approach by computing query cover rate, i.e., the proportion
of queries resolved by using views. When the storage space constraint is hard
($S = 0.05\% \times S_V$), average cover rate is already 23\%. It reaches 100\%
when the storage space constraint is relaxed.

\subsection{Index selection results}
\label{sec:IndexSelection}

\begin{figure}[hbt]
    \centering
    \includegraphics[width=9cm]{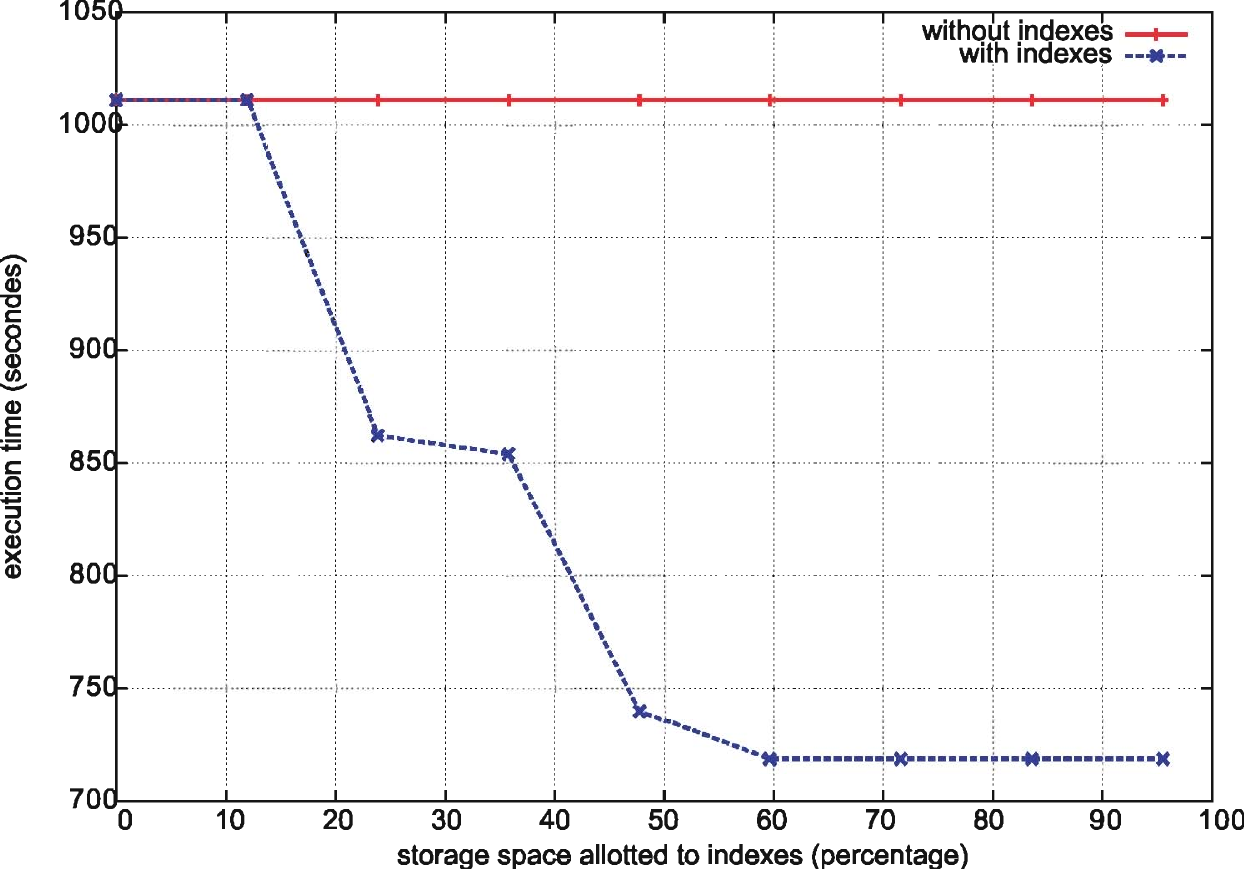}
    \caption{Indexing experiment results}
    \label{fig:exp-i}
\end{figure}

In these experiments, we have fixed the minimal support parameterized in the Close
frequent itemset mining algorithm to 1\%. This value gives the highest number
of frequent itemsets and consequently the highest number of candidate indexes.
This helps vary storage space $S$ within a wide interval. We have measured
query execution time according to the percentage of storage space allotted for
indexes. This percentage is computed from the space $S_I$ occupied by all
indexes. Figure~\ref{fig:exp-i} shows that execution time decreases when
storage space occupation increases. This is predictable because we create more
indexes and thus better improve execution time. We also observe that the
maximal time gain is about 30\% and it is reached for space occupation $S =
59.64\% \times S_I$.

Finally, these experiments also showed that our index selection strategy helped
select a portion of candidate indexes that allows to achieve roughly the same
performances than the whole set of candidate indexes. This guarantees
substantial gains in storage space (40\% on an average) and decreases index
maintenance cost.

\subsection{Joint index and materialized view selection results}
\label{sec:JointIndexAndMaterializedViewSelection}

\begin{figure}[hbt]
    \centering
    \includegraphics[width=12cm]{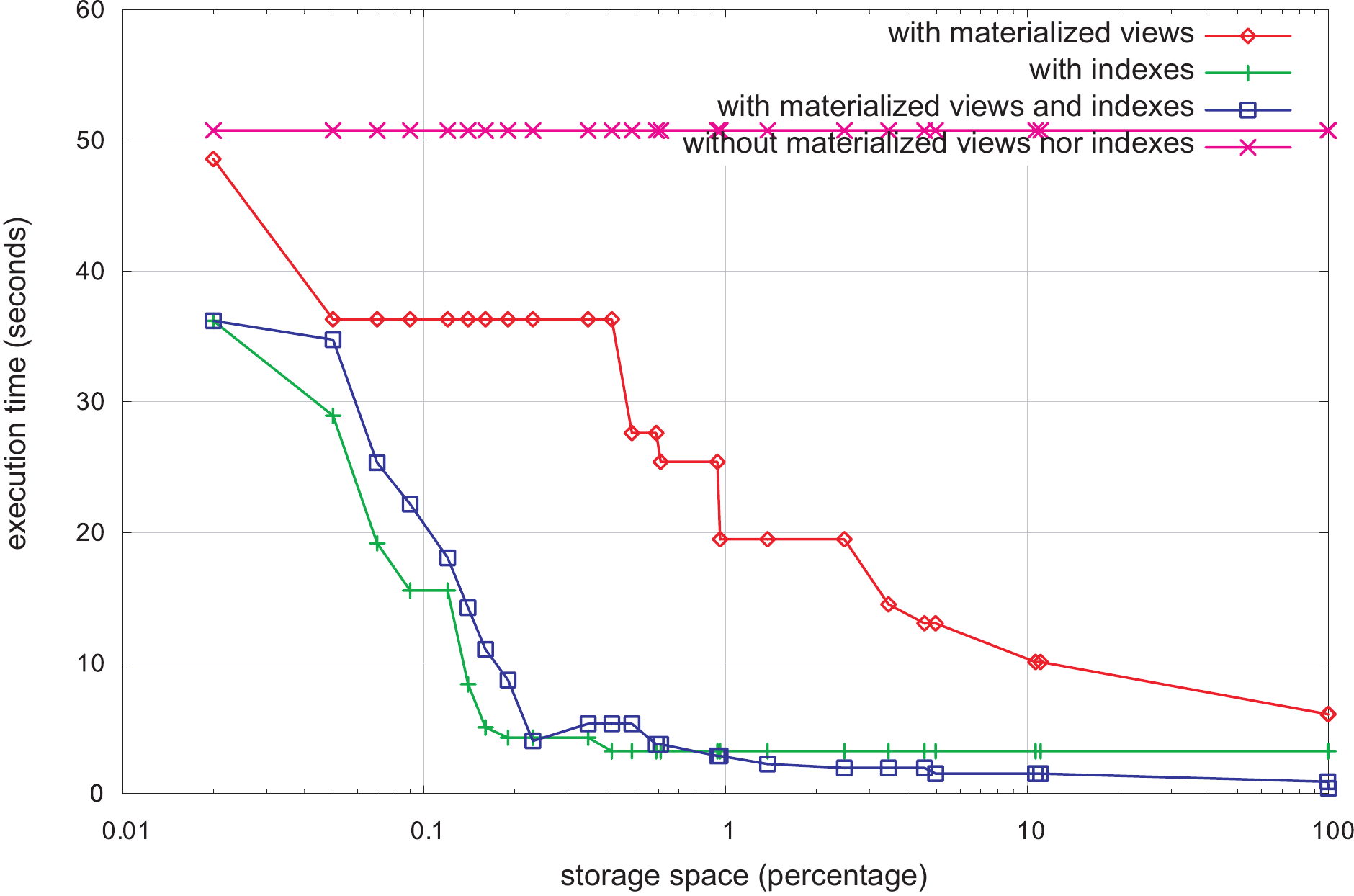}
    \caption{Joint view materialization and indexing experiment results}
    \label{fig:resultat_vues_index}
\end{figure}

Eventually, we have compared the efficiency of isolate materialized view
selection, isolate index selection and joint materialized view and index
selection. We have measured query execution time in the following cases:
without materialized views nor indexes (reference plot), with materialized
views only, with indexes only and with both materialized views and indexes
(simultaneously selected). Figure~\ref{fig:resultat_vues_index} represents the
variation of response time with respect to the storage space $S$ allotted to
materialized views and indexes. $S$ is expressed in percentage of total space
$S_{VI}$ occupied by all indexes and materialized views, achieved when we apply
our strategy without any storage space constraint. Note that we used a
logarithmic scale on the X axis to better visualize the results.

Figure~\ref{fig:resultat_vues_index} shows that jointly selecting materialized
views and indexes allows better performance than selecting indexes or views
separately when storage space is large. However, when it is small, isolate
index selection is more competitive than the other solutions. This may be
explained by the fact that index size is generally significantly smaller than
materialized view size. Then, we can store many more indexes than materialized
views in a small space and achieve a better performance. In conclusion, indexes
should thus be privileged when storage space is strongly constrained.

\section{Conclusion and perspectives}
\label{sec:ConclusionAndPerspectives}

We have presented in this paper an approach for automatic data warehouse
performance optimization. Our main contribution in this field mainly relates to
exploiting knowledge about the data warehouse and its usage. Knowledge may
either be formalized expertise, or automatically extracted with the help of
data mining techniques. This approach allowed us to reduce the dimensionality
of the materialized view and index selection problem, by proposing a reduced
and pertinent candidate object configuration. We also have explicitly taken
view-index interactions into account, to propose a final object configuration
that is as close as possible to the optimum.

We have designed our approach to be generic. Data mining techniques and cost
models we exploit are indeed not related to any system in particular. They may
be applied on any host DBMS. Our materialized view and index strategies are
also modular: each step (candidate object selection, cost computation...)
exploits interchangeable tools. The data mining techniques and cost models we
used could easily be replaced by others. Moreover, we could also extend our
approach to other performance optimization techniques, such as buffering,
physical clustering or partitioning~\cite{agrawal04,zilio04,bel05evo}.

Though we have systematically tried to demonstrate the efficiency of our
proposals by experimenting on real-life systems such as Oracle, up to now, we
have not been able to compare our proposals to existing approaches \emph{in
situ}. Those that are developed by DMBS vendors
\cite{agrawal04,dageville04,zilio04} necessitate the acquisition of the
corresponding system. Furthermore, they are implemented as ``black boxes'' that
are often hard to tinker with. Finally, research proposals from the literature
are not always available as source or executable code and, when it is the case,
they operate in one given environment and must often be reimplemented. We shall
have to get over these difficulties to complete our solutions' experimental
validation, though.

Finally, the main possible evolution for our work resides in improving our
solutions' automaticity. We indeed perform static performance optimization. If
the input query workload significantly evolves, we must rerun the whole process
to preserve performance. Dynamic materialized view selection approaches that
have been proposed to optimize refreshing times \cite{kot99dyn,shah06} are more
efficient than static approaches. We must work in this direction for optimizing
query response time.

Our main lead is to exploit our approach's modularity by replacing the data
mining techniques we used by incremental frequent itemset mining \cite{leung05}
or clustering \cite{jai99data} techniques. Studies related to session detection
that are based on entropy computation \cite{qui05mac} could also be exploited
to detect when to rerun the (incremental) selection of materialized views and
indexes.

\singlespacing

\bibliographystyle{theapa}
\bibliography{biblio}

\end{document}